# Convolutional XGBoost (C-XGBOOST) Model for Brain Tumor Detection

Muyiwa Babayomi
Faculty of Computer and Informatics
Bournemouth University
*Dorset, United Kingdom*
s5436304@bournemouth.ac.uk

Oluwatosin Atinuke Olagbaju
Faculty of Computer and Informatics
Bournemouth University
*Dorset, United Kingdom*
s5432408@bournemouth.ac.uk

Abdulrasheed Adedolapo Kadiri
Faculty of Computer and Informatics
Bournemouth University
*Dorset, United Kingdom*
s5436150@bournemouth.ac.uk

*Abstract*—Brain tumors are masses or abnormal growths of cells within the brain or the central spinal canal with symptoms such as headaches, seizures, weakness or numbness in the arms or legs, changes in personality or behaviour, nausea, vomiting, vision or hearing problems and dizziness. Conventional diagnosis of brain tumour involves some tests and procedure which may include the consideration of medical history, physical examination, imaging tests (such as CT or MRI scans), and biopsy (removal and examination of a small piece of the tumor tissue). These procedures, while effective, are mentally strenuous and time demanding due to the manual examination of the brain scans and the thorough evaluation of test results. It has been established in lots of medical research that brain tumours diagnosed and treated early generally tends to have a better prognosis. Deep learning techniques have evolved over the years and have demonstrated impressive and faster outcomes in the classification of brain tumours in medical imaging, with very little to no human interference. This study proposes a model for the early detection of brain tumours using a combination of convolutional neural networks (CNNs) and extreme gradient boosting (XGBoost). The proposed model, named C-XGBoost has a lower model complexity compared to purely CNNs, making it easier to train and less prone to overfitting. It is also better able to handle imbalanced and unstructured data, which are common issues in real-world medical image classification tasks.

To evaluate the effectiveness of the proposed model, we employed a dataset of brain MRI images with and without tumours. The dataset used for training and testing was retrieved from figshare public repository, it contains 2 folders with 253 positive and negative brain MRI Images. The data was pre-processed and augmented to ensure a diverse and more representative sample. The C-XGBoost model was trained and validated on the dataset, and the results were compared to those of a non-hybrid CNN-based model. The model extracts data features using the DenseNet-121 transfer learning model. The proposed model achieved an F1 score of 0.97 and an accuracy of 99.02%, outperforming the CNN-based model, which achieved an accuracy of 98.8%, demonstrating its potential as a more reliable technique for detecting brain tumours. The C-XGBoost model had a lower training and validation loss, indicating better generalization to the test set. Our experimental results showed that the proposed model achieved high levels of accuracy in detecting brain tumours from medical images, making it a viable approach for early detection of brain tumours.

## I. INTRODUCTION

According to data from Cancer Research UK, in 2018, there were an estimated 10,800 new cases of brain and central nervous system tumors diagnosed in the United Kingdom [1]. This number includes both benign (non-cancerous) and malignant (cancerous) tumors. The incidence rate of brain tumors in the UK varies by age, with the highest rates occurring in children under the age of 15 and in adults over the age of 65. The most common type of brain tumors in the UK is a glioma, which accounts for around half of all brain tumors. WHO estimates that brain and other central nervous system tumors accounted for around 2.2% of all deaths from cancer in 2018, which was approximately 784,000 deaths worldwide. This represents a relatively high mortality rate, as brain tumors are often difficult to treat and can be aggressive in nature.

Brain tumors can be caused by a variety of factors, including genetics, radiation exposure, and certain viral infections. However, in many cases, the cause of a brain tumor is unknown. Treatment for brain tumors may include surgery, radiation therapy, chemotherapy, or a combination of these approaches. The specific treatment plan will depend on the type and stage of the tumor, as well as the patient's age and overall health. The prognosis for brain tumor patients varies widely, depending on the type and stage of the tumor, as well as the patient's age and overall health. Some brain tumors can be cured with surgery or other treatments, while others are more difficult to treat and may be incurable.

Early detection of brain tumors can potentially improve the chances of successful treatment and increase the survival rate of patients diagnosed with the disease. Machine learning techniques, such as ConvXGB, can be used to identify tumors at an earlier stage when they are more likely to be successfully treated. The CNN component of our proposed model extracts distinctive features from medical images, while the xgboost component uses these features to make predictions about the presence of tumours. The model was trained on a dataset of medical brain scan images and was able to detect different types of brain tumours. In addition, the model was able to handle a wide variety of image types and conditions, making it highly versatile and applicable to a wide range of medical scenarios. The results of this research demonstrate the potential of machine learning algorithms for early and accurate detection of brain tumours, paving the way for improved patient outcomes and reduced healthcare costs.

## II. LITERATURE

There is no doubt that machine learning has become a crucial field in the biomedical sciences because it provides techniques for analysing high-dimensional and multimodal data. However, current approaches have difficulty incorporating structural and functional imagery, as well as genomic, proteomic, and ancillary data. As far as learning machines are concerned, the human brain is by far the most flexible and powerful. It is for this reason that the machine learning community has become increasingly interested in neuroscience, seeking to understand brain-based learning systems as well as identify new theories and architectures.

Machine learning-aided diagnosis aims to assist doctors in disease diagnosis using artificial intelligence (AI) and deep learning techniques. One of such technique is Convolutional neural networks, which can be used to analyse and evaluate medical images to detect and extract abnormal features or lesions. Qiao Z. et al - examined the use of CT image characteristics in combination with blood tumour markers and a convolutional neural network (CNN) algorithm to diagnose pancreatic cancer [2]. The model uses machine learning algorithms, tensor flow, and neural networks to analyse various related health parameters which are parsed as input variables through the fully connected layers of the CNN model, to predict the likelihood of the presence of the diseases. The study also examined the use of CT image characteristics in combination with blood tumour markers and the three-dimensional (3D) convolution neural network (CNN) algorithm to make a clinical diagnosis of pancreatic cancer [2]. Neural networks are known to excel at prediction problems involving unstructured data such as images [3]. However, numerous disease processes in patients present a significant barrier in segmentation, radionics and various detection methods that are relevant for early treatment and prognosis.

Machine learning has made significant advances in the medical field by providing tools and techniques for analysing complex data. By applying machine learning algorithms, such as decision trees, random forests, and naive Bayes, to medical records, researchers can assess the performance of these algorithms using various evaluation metrics. The accuracy of these algorithms can sometimes reach up to accuracy levels that medical experts routinely operate at. In this study [3], researchers compared the performance of various supervised machine learning techniques for the prediction of Alzheimer's disease. To ensure a fair comparison of performance analysis, they selected studies that used multiple machine learning algorithms on the same dataset. They established a common benchmark for datasets and scope to account for the wide variability in clinical data and research scope. This research showed that the algorithms being used were effective in detecting diseases in their early stages with a high degree of accuracy.

*A. Convolution Neutral Network (CNN) In AI*

Convolutional neural networks (CNNs) are a state-of-the-art development in artificial intelligence. CNNs were introduced by Yann LeCun as a new generation of technology for image processing, replacing older methods [4]. CNNs are a variant of regular neural networks that process images by taking in a 3D input volume and transforming it into a 3D output volume. CNNs consist of several layers that learn directly from images and transform them to produce output. In this research, the CNN model was used to perform image analysis tasks such as object recognition, image classification and segmentation. To predict and detect diseases in patients, the CNN model uses a key concept called shared weights and biases.

Shared weights and biases in CNNs refer to the process of training the model to detect the same features in different parts of an image, allowing the network to recognize objects in an image [5] [6]. This reduces the number of parameters the network needs to predict and detect images by defining kernels or filters in a convolutional layer based on various feature maps. CNNs are a powerful tool for image analysis and can be used to detect and predict diseases.

In the research paper [7], CNN is represented as a mathematical theory. Assume the input equation as:

$$p = \{x(m,n) | 1 \leq m \leq W, 1 \leq n \leq H\}, \quad (1)$$

$x(m,n)$ Is the intensity of the pixel in the kernels $(w_k \times h_k)$, $k$ is the convolution which produces the map $y$ from the image and $p$ is apply to filter $k$. $S_k$ are the filters that are slid through the image by $k$. This is defined as discrete convolution which is show in the equation below:

$$(p \oplus k)_{m,n} = \Sigma_{w=-wr}^{hh} \Sigma_{v=-hk} k_{w_v} I_{n+y+v}, \quad (2)$$

The convolution operation is applied to the input along with the additive bias indexed by $f\varepsilon_{\{1,\ldots 1f(i)\}}$ so the output $y$ is derived from the previous layer:

$$y_i^{(i)} = \phi \left( B_i^{(i)} + \Sigma_{j=1}^{f(i-1)^1} \right)_1 k * y_j^{(i-1)}, \quad (3)$$

- $\phi$ = rectified linear unit (ReLU).
- $B_i^{(l)}$ = Bais Matrix.
- $k$ = this is the filter of size $2wk \times 2nk + 1$

This is the output of the layer, $l$, $i$, $y_i^{(i)}$, position of $(m,n)$ is:

$$\left(y_i^{(U)}\right)_{m,n} = \phi \left( \left(B_i^{(l)}\right)_1 \right)_{m,n} + \Sigma_{j-1} f^{(i-1)} (k_{i,j}^{(l)} \otimes y_j^{(i-1)})_{m,n}, \quad (4)$$

This output with the highest valuable within the neighbourhood is replaced by the output of the previous layer. The previous output in the layer $\left(y_i^{(U)}\right)_{m,n}$ shows that $\emptyset$ activation function with $p(.)$ is the pooling process that stride $S_p$, and $v_p \times w_P$. There are two types of pooling in this operation which are medium pooling and maximum pooling. The maximum pooling is the max value in the window that are taken. The maximum pooling output process function is made below:

$$p\left(y_i^{(U)}\right)_{m,n} = max\left(y_i^{(U)}\right)_{m,n} \quad (5)$$

If the max function in this operation is appliable to the maximum pooling process in each window dimension, then the operation will become:

$$dP\left(y_i^{(U)}\right)_{m,n} = \left((w - wk) / S_P + Px\right) * \left((H - |h_k|)/S_p + 1\right), \quad (6)$$

The final layer of the CNN model is called the Fully Connected (FC) layer, and it takes the output from the previous pooling layer as its input. In the FC layer, all the neurons in the preceding layer are connected to the following layer. We apply a multilayer perception equation that includes L FC layers, where f_1^((L)) is the numerical value representing the feature size of the maps.

$$\left(y_i^{(U)}\right)_{m,n} = \phi\left(\Sigma_{p=1}^{f1^{(l-1)}} \Sigma_{q=1}^{f2^{(l-1)}} \Sigma_{r=1}^{f3^{(l-1)}} \omega_{i,p,q,r}^{(l)} \left(y_j^{(l-1)}\right)_{q,r}\right). \quad (7)$$

This operation above explains $\omega_{i,p,q,r}^{(l)}$ is the connection of the weight unit at $(m, n)$ in the maps inside $i$, layer as $l$ in position of $(q, r)$ at the feature map, $p$ and $(l-1)$.

*B. CNN Model Architecture*

In a convolutional neural network (CNN), there are three main types of layers: input, convolutional, and fully connected layer. The input layer receives the input data, which may be an image, text, or numerical data. The convolutional layers use filters to identify important features in the input data. The features extracted by the convolutional layers are then processed by one or more fully connected layers, which utilize them to make predictions, classification or perform other tasks. The accuracy of the CNN model depends on the ability of the convolutional layers to effectively learn and extract meaningful features from the input data.

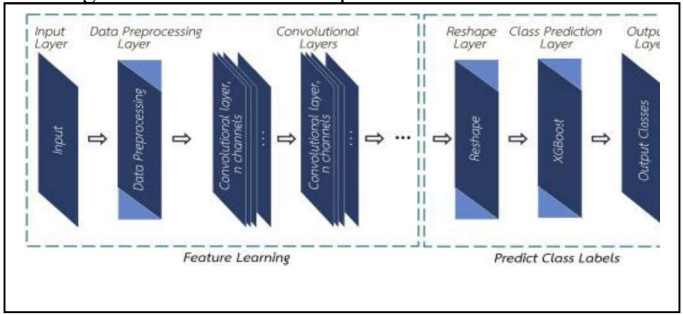

Fig. 1 CNN model

*C. Feature Learning*

Data pre-processing: Normalization and standardization can be applied to the input data in a machine learning model. Normalization refers to the process of scaling the input data to have a range of values between 0 and 1, while standardization refers to the process of scaling the random variable of the input data by subtracting the mean of the data from each data point and then dividing the result by the standard deviation of the data in order to ascertain a mean of zero and a standard deviation of one.

Input layer: The input layer is the first layer in a convolutional neural network (CNN) and it receives the raw input data. In the context of the proposed model C-XGBoost, the input data may be a dataset of tuples $(x_i, y_i)$ where $i$ is the index in the dataset. $x_i$ is defined as the feature matrix and $y_i$ represent the class label to $x_i$. The input layer is responsible for providing the raw input data to the rest of the model.

It is important to note that normalization and standardization are not always necessary or appropriate for all types of input data or models. In some cases, these pre-processing steps may improve the performance of the model, while in other cases they may have little or no effect. It is usually a good idea to experiment with different pre-processing techniques and evaluate their effect on the model's performance.

Convolutional layer: The convolutional layer is responsible for applying convolutional filters to the input data to extract features from it. In C-XGBoost, the convolutional layer is used to extract features from the input data to feed into the XGBoost layer, which learns a decision tree model from the data. The output of the convolutional layer is typically a tensor with dimensions $\left(\sqrt{N}, \sqrt{N}, z^{(l)}\right)$ using the tensor in its dimension, where z^((l)) is the number of filters applied in the $l$-th layer.

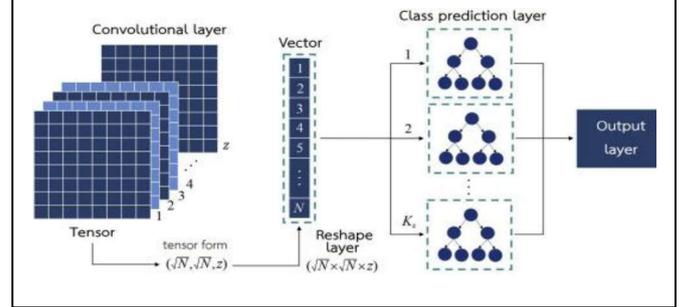

Fig. 2 CNN layer architecture.

The number and complexity of the layers in a convolutional neural network (CNN) should be carefully chosen to balance the model's ability to learn complex patterns in the data with the risk of overfitting. Overfitting occurs when a model is trained on a small or insufficiently diverse dataset, or when the model is too complex for the underlying data. To avoid overfitting, it is often necessary to use techniques such as regularization or early stopping to reduce the complexity of the model and consequently improve its generalization performance.

In the case of a CNN designed to predict and detect diseases, it may often be beneficial to use a deeper network with more layers in order to capture more complex patterns in the data. However, it is important to carefully evaluate the performance of the model on a validation set and tune the number and complexity of the layers to achieve the best balance between accuracy and overfitting.

*D. Class Label Prediction*

In this part of a convolutional neural network (CNN), there are typically three types of layers that work together to predict class labels from input data: reshape layers, fully connected layers, and output layers.

Reshape layer: This layer is used to change the shape or dimensions of the input data. A reshape layer may be used to convert the output of the convolutional layers from a tensor with three dimensions (e.g., width, height, and depth) to a tensor with two dimensions (e.g., a vector). This is often done to prepare the input data for the fully connected layer.

Fully connected layer: In this layer, all the neurons in one layer are connected to all the neurons in the next layer. In a CNN designed to predict class labels, the fully connected layer(s) may be used to learn more complex relationships between the features extracted by the convolutional layers and the target classes. The class prediction layer in the statement provided is likely referring to one or more fully connected layers.

Output layer: The output layer is the final layer in a CNN, and it is responsible for producing the final prediction or output of the model. In a classification task, the output layer may have as many neurons as there are classes in the dataset, and each neuron corresponds to a different class. The output layer may use a variety of activation functions, such as SoftMax, to produce a probability distribution over the classes.

$$\tilde{L}^{(t)}(q) = -\frac{1}{2} \Sigma_{j=1}T \frac{(\Sigma_{i \in I}, g_i)^2}{\Sigma_{i \in I}, h_i + \lambda} + \gamma T$$

*E. Extreme Gradient Boosting (Xgboost) In AI*

Jason et al proposed the classification of data features using extreme gradient boosting, which is a supervised learning algorithm that can be used to accurately predict various diseases by combining several weaker models [8,9]. This method involves ensembling K classification and regression trees, each with K nodes. The prediction is the sum of the scores of each tree in the K-node:

$$\hat{y}i = \varphi(x_i) = \Sigma_{k-1}^k f_k(x_i) F_k \in f_j$$

This operation states that $x_i$ is defined as the members of the training set, $\hat{y}i$ is set of the equivalent class title in the dataset, $F_k$ is the total value for the given K tree and $F$ is the set that contain the number in K scores for all the K classification and regression tree. To improve this operation, we added regularization process to it.

$$\mathcal{L}(\varphi) = \Sigma_i l(\hat{y}_i, y_i) + \Sigma_k \Omega(f_k)$$

The first operation shows that $l$ is the missing function that measure the total variance between the key value $y_i$ and the prediction $\hat{y}_i$ and $\Omega$ represents the importance result of the model:

$$\Omega(f) = \sigma T + \frac{1}{2} \lambda \sum_{j=1}^{T} w_j^2$$

The second operation shows that both $\sigma$ and $\lambda$ are the constants that controls the learning algorithm degree, $T$ is defined as the number of leaves in the tree and $w$ is the measured weight of each leaf of the tree. Gradient boosting has great effect in both regression and classification problems. They are used with the missing functions for extending the constant terms reduction to give a simpler objective term:

$$\tilde{L}^{(t)} = \Sigma_{i-1} \left[ g_i f, (x_i) + \frac{1}{2} h_j f_1^2(x_1) \right] + \Omega(f_t)$$

Where $I_j = \{i | q(x_i) = j\}$ indicate the data set of leaf $t$, and $g_i = \frac{\partial i(\hat{y}^{(t-1)}, y_i)}{\partial \hat{y}_i^{(t-1)}}$, $h_i = \frac{\partial^2 i(\hat{y}^{(t-1)}, y_i)}{\partial (\hat{y}_i^{(t-1)})^2}$

The $g_i$ and $h_i$ are the total image statistics of the missing function. The perfect weight of the leaf in the K tree, the condition of the tree, for a given tree can be obtained in below operation:

$$\omega_j^x = -\frac{\Sigma_{i \in I}, g_i}{\Sigma_{i \in I}, h_i + \lambda}$$

$$\tilde{L}^{(t)}(q) = -\frac{1}{2} \Sigma_{j=1} T \frac{(\Sigma_{i \in I}, g_i)^2}{\Sigma_{i \in I}, h_i + \lambda} + \gamma T$$

$\omega_j^x$ is the perfect weight given in the leaf.
$j$ is the leaf of the tree.
$q$ is the condition of a tree structure.
$q(x_i)$ is the tree structure that is given.

In this operation below, we are showing the evaluation of a splitting candidate score in the total datasets of the right and left nodes then it was occurred to loss reduction in its process:

$$\mathcal{L}_{split} = \frac{1}{2} \left[ \frac{(\Sigma_{i \in IL}, g_i)^2}{\Sigma_{i \in IL}, h_i + \lambda} + \frac{(\Sigma_{i \in IR}, g_i)^2}{\Sigma_{i \in IR}, h_i + \lambda} + \frac{(\Sigma_{i \in I}, g_i)^2}{\Sigma_{i \in I}, h_i + \lambda} \right] - \gamma$$

*F. C-XGBoost Machine Learning Technique*

Convolutional Extreme Gradient Boosting (C-XGBoost) is a variant of the Extreme Gradient Boosting (XGBoost) algorithm, which is a popular and efficient implementation of the Gradient Boosting algorithm for machine learning. Like the original XGBoost algorithm, C-XGBoost is designed for efficient training of decision tree models for regression, classification, and ranking tasks. However, C-XGBoost extends the capabilities of XGBoost by incorporating convolutional neural network (CNN) layers into the model. This allows C-XGBoost to handle input data that has a grid-like structure, such as images, and to learn more complex patterns from the data. They are utilized in a wide range of machine learning applications like NLP and genomics.

The ConvXGB algorithm combines the strengths of both CNNs and XGBoost to reduce model complexity and the number of parameters required for prediction. This is achieved by using CNNs without pooling or fully connected layers, and by using XGBoost as the final layer. This reduces the risk of overfitting and makes the model more efficient and easier to train. The C-XGBoost algorithm is a powerful tool for image analysis and disease prediction, and it has the potential to greatly improve the accuracy and efficiency of these tasks.

The individual contributions of CNN and XGBoost in the proposed C-XGBoost model are as follows:

Convolutional neural networks have several layers of convolutional, pooling, and fully connected layers. Convolutional layers apply a set of filters to the input data, which are used to extract features from the data. Pooling layers reduce the dimensionality of the data by applying a function such as max pooling or average pooling to subregions of the data. Fully connected layers connect all the neurons in one layer to all the neurons in the next layer, allowing the network to learn more complex relationships between the features.

The XGBoost algorithm is used to learn a decision tree model from the input data, and the convolutional layers are used to identify features from the data to feed into the XGBoost model.

*G. C-XGBOOST MODEL APPROACH*

Given a C-XGBoost model trained on a dataset of sample size M, where each element $(x_j, y_j)$ Consists of a vector of tree numbers $x_j$ in $R^N$ and a corresponding vector $y_j$, The model can be implemented following this approach:

Importing Library: We utilize python libraries such as pandas, matplotlib, Kera, NumPy, seaborn, TensorFlow, CNN and cv2 to import, analyse and process our data before the model is trained on the dataset. The proposed brain tumour prediction technique identifies the likelihood of a patient having a brain tumour from the MRI scan of the brain. The model is trained to classify the presence or absence of brain tumours based on observable imaging features.

Loading of Data: The proposed machine learning model is trained on a dataset sourced from public repositories. The dataset used for the model is sourced from figshare public repository.

Exploratory Data Analysis: A cross-classified analysis is carried out on the dataset using graphical and non-graphical techniques. This helps to collect, clean, pre-process and visualize data to have insight into the data features and types. This step is crucial for correcting missing data before pre-processing them into comprehensible formats for further analysis. This step involves image normalization and standardisation. Another crucial benefit of this step is to remove biases from skewed data distributions.

Prepare the input data: It is important to pre-process the input data to ensure that it is in a suitable format for input into the C-XGBoost model. This may involve normalizing or standardizing the data as stated in step three above, converting it to a tensor with appropriate dimensions, and splitting the data into a training set and a test set. The training set is a subset of the data that is used to train the model, while the test set is a separate subset of the data that is used to evaluate the performance of the trained model.

Define the CNN architecture: The architecture of the CNN in a C-XGBoost model plays a critical role in its performance. This involves choosing the number and type of layers to use, as well as the hyperparameters for each layer. For example, the number of filters and kernel size in a convolutional layer can be chosen to extract relevant features from the input data.

1. Initialize the weights and biases: In a CNN, the weights and biases of the layers are typically initialized randomly using a Gaussian distribution. The mean and standard deviation of the distribution can be chosen based on the characteristics of the data and the desired properties of the model.

2. Feed the input data through the CNN: Given a set of input data X and a set of corresponding labels Y, the input data can be fed through the CNN using the following equation:

   $Z = f(W*X + b)$

   where Z is the output of the CNN, f is an activation function, W is the weight matrix for the layer, and b is the bias vector.

3. Calculate the loss function: The loss function is used to measure the discrepancy between the predicted output of the CNN and the true labels. This can be done using a variety of loss functions, such as the cross-entropy loss:

   $L = -\Sigma_i Y\_i * \log(\hat{y}\_i) + (1 - Y\_i) * \log(1 - \hat{y}\_i)$

4. Backpropagate the error: After calculating the loss, the error can be backpropagated through the CNN using the following equation:

   $\partial L/\partial W = (\hat{y} - Y) * f'(Z) * X$
   $\partial L/\partial b = (\hat{y} - Y) * f'(Z)$

5. where $\partial L/\partial W$ is the gradient of the loss function with respect to the weights, $\partial L/\partial b$ is the gradient with respect to the biases, and $f'(Z)$ is the derivative of the activation function with respect to the output of the CNN.

6. Update the weights and biases: After backpropagating the error, the weights and biases can be updated using an optimization algorithm, such as stochastic gradient descent, using the following equation:

   $W\_new = W - \eta * \partial L/\partial W$
   $b\_new = b - \eta * \partial L/\partial b$

   where $\eta$ is the learning rate and W_new and b_new are the updated weights and biases, respectively.

7. Extract features from the CNN: After the CNN has been trained, the output of the convolutional layers can be extracted as features for use in the XGBoost model. These features can be fed into the XGBoost model along with the target labels to train the model.

8. Train the XGBoost model: The XGBoost model can be trained using the gradient boosting algorithm to minimize the loss function and improve prediction accuracy. This involves building a series of decision trees and adding them together to form a strong prediction model.

9. Evaluate the model: After training the ConvXGB model, it is important to evaluate its performance on a separate test dataset to assess its generalization ability and determine its prediction accuracy. This can be done by comparing the predicted labels to the true labels and calculating evaluation metrics such as accuracy, precision, and recall.

*H. Existing Disease Prediction Models*

There has been significant research using machine learning techniques to detect and diagnose various diseases, some of which have achieved notable success. For example, Ross et al. (2020) used machine learning to identify a devastating neurological disorder at various diagnostic stages, for which there is currently no known cure [10]. Perez et al. proposed a method that uses a neural network to analyse DNA CpG methylation data in order to improve the early detection of Huntington's Disease (HD) [11]. Zhou et al. (2021) utilized artificial intelligence and computer vision to segment the sublingual vein region in order to improve the non-invasive identification of multiple diseases [12]. Table 1.3 summarizes the performance of existing machine learning models that have been used to predict various diseases.

Table 1. performance characteristics of existing machine learning models for various diseases.

| Reference | Study Title | Machine Learning Technique Used | Accuracy |
|---|---|---|---|
| Koo et al. [13] | Deep learning-based diagnosis of breast cancer using mammography images | Deep Convolutional Neural Network | 95.50% |
| Ezzat et al. [14] | Predicting the onset and progression of Alzheimer's disease using machine learning | Convolutional Neural Network | 95.00% |
| George et al. [15] | Predicting diabetes using logistic regression | Logistic Regression | 84.70% |
| Mavroforakis et al. [16] | A machine learning approach for early diagnosis of Parkinson's disease using non-invasive biomarkers | Convolutional Neural Network | 95.60% |
| Imran et al. [17] | A deep learning approach for predicting the severity of diabetic retinopathy | Deep Convolutional Neural Network | 92.70% |
| Pechenizkiy et al. [18] | Predicting heart failure using decision tree and random forest algorithms | Decision Tree, Random Forest | 76.70% |
| Gao et al. [19] | Predicting the likelihood of stroke using a hybrid machine learning approach | Decision Tree, Artificial Neural Network | 80.60% |
| Wang et al. [20] | Predicting the likelihood of breast cancer using support vector machine | Support Vector Machine | 81.30% |

## III. METHODOLOGY

The methodology proposed in this research investigates the effectiveness of convolutional extreme gradient boosting (c-XGBoost) method for training a dataset of brain MRI images with and without tumors. To achieve this, we pre-processed the dataset by applying appropriate image augmentation techniques and normalization to ensure that the input data was in a suitable format for the C-XGBoost model. Next, we extracted features from the pre-processed images using a convolutional neural network (CNN) model, specifically the DenseNet121 architecture. The extracted features were then fed into the c-XGBoost model for training and evaluation. We evaluated the performance of the c-XGBoost model using standard metrics such as accuracy and F1 score and compared its performance to other commonly used machine learning algorithms. Our results showed that the c-XGBoost model achieved superior performance in predicting the presence of brain tumours in the dataset. The proposed model utilizes the CCO (creative common) licensed figshare dataset containing 253 images of human brain MRI images which are classified into 3 classes: glioma - meningioma - and pituitary.

Here are a few key advantages of using a convolutional extreme gradient boosting (C-XGBoost) approach over a non-hybrid convolutional neural network (CNN) approach:

- C-XGBoost combines the ability of CNNs to extract features and patterns from images with the ability of XGBoost to effectively model and make predictions using those features. This combination allows for the model to not only effectively identify relevant features in the data, but also to make accurate predictions using those features. Additionally, XGBoost can handle large and complex datasets, as well as handle missing or incomplete data, which can be beneficial in the medical field where data may be limited or unreliable. Overall, the use of C-XGBoost in medical image analysis can lead to improved performance and accuracy compared to the use of a single method, such as a CNN, on its own.

- The lower model complexity of C-XGBoost is achieved using XGBoost, a gradient boosting algorithm, which combines the predictions of multiple weak models, rather than relying on a single complex model. This ensemble approach results in a more robust model that is less likely to overfit to the training data, and thus, is more likely to generalize well to unseen data. Additionally, the use of XGBoost can also help to reduce the risk of overfitting by implementing regularization techniques such as early stopping and tree pruning. Overall, the lower model complexity of C-XGBoost makes it easier to train and more reliable for use in real-world applications.

- While CNNs can extract feature maps from images and identify patterns, they are not as interpretable as traditional machine learning algorithms. On the other hand, XGBoost is a powerful and interpretable decision tree-based algorithm that is widely used in machine learning. By combining these two approaches, C-XGBoost can retain the ability to extract features from images and make accurate predictions, while also providing a level of interpretability and explainability that may be important in a medical setting. This can be particularly useful when it comes to understanding why a particular diagnosis or treatment was recommended, and how certain features in the images were used to arrive at that decision.

- C-XGBoost can handle imbalanced datasets better than CNNs because it uses gradient boosting, which focuses on minimizing the misclassification error rather than the overall error. This means that C-XGBoost will give more weight to the minority class and try to classify those samples correctly. In medical image classification tasks, where the number of brain tumour samples belonging to the minority class is often much smaller than the number of samples without brain tumours belonging to the majority class, this can lead to improved performance and accuracy.

- C-XGBoost can handle missing values and noisy data better than CNNs, which can be an issue in real-world medical datasets. Noisy data can negatively impact the performance of a machine learning model. In the context of medical image classification, noisy data can come from various sources, such as errors in data collection or annotation, or variations in image quality or resolution. C-XGBoost is less sensitive to such noise compared to CNNs, as it uses a combination of feature maps extracted from CNNs and the XGBoost algorithm to make predictions. The XGBoost algorithm can handle noise and missing values in the data by using decision trees, which can be robust to such issues and make more accurate predictions.

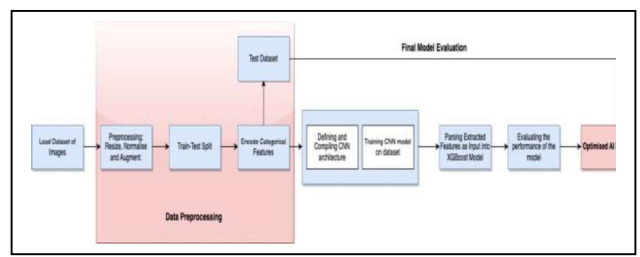

Fig. 3 Block Diagram of Proposed C-XGBoost Model.

The C-XGBOOST algorithm for detecting brain tumours in python is demonstrated showing results from the code including evaluation metrics for analysing the algorithm's performance. The implementation of pooling layers and the output of the algorithm are also shown.

*A. Importing Required Library*

The necessary python libraries were imported which include NumPy, pandas, TensorFlow, pathlib, matplotlib.pyplot, pymatreader, sklearn, keras and xgboost. All of which are tools and functions essential for implementing and training the proposed C-XGBOOST model. Shown below is the screenshot of the library importation code as implemented in Jupyter notebook.

*B. Loading dataset*

A dataset of medical brain scan images that includes both examples of brain tumor presence and healthy brain scans, was collected from the figshare public repository, unzipped and loaded into the model. The code snippet shown in fig 3.3 downloads a zip file from the specified URL and saves it to the local system with the specified file name "brain_tumor_dataset.zip". The "wget" command is used to download files from the source, and the "-O" flag specifies the name to be used for the saved file. The URL provided in the command points to a file on figshare.com, a public online scientific repository which contains the utilized dataset. The unzipped files are in .mat format. The already imported python library read_mat reads data in MATLAB format.

Fig. 4 Dataset loading from figshare repository

Fig. 4 Data unzipping

*C. Data pre-processing*

Data pre-processing is a crucial step in the model training process, as it helps to ensure that the data is in a suitable format for further analysis and modelling. Some common steps involved in data pre-processing include but not limited to:

Data cleaning: This involves removing or correcting any inaccuracies, inconsistencies, or missing data in the dataset.

Data integration: This involves combining data from different sources or formats to create a single, unified dataset.

Data transformation: This involves transforming the data into a suitable format for further analysis, such as scaling or standardizing numerical variables or encoding categorical variables.

Data reduction: This involves selecting a subset of the data that is relevant for the analysis and discarding the rest.

Data discretization: This involves converting continuous numerical variables into discrete bins or categories.

Data visualization: This involves creating charts, graphs, or other visualizations to help understand the data and identify trends or patterns. Shown in fig 4 are some loaded dataset images snipped from the Jupyter notebook IDE.

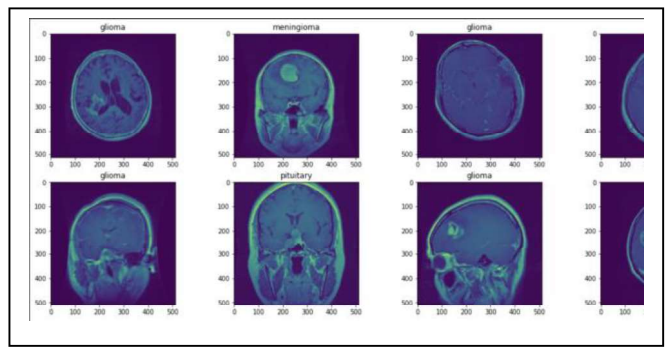

Fig. 5 some examples of visualised image features.

*D. Collecting Feature labels*

In the proposed model, the collect_features_labels function is responsible for collecting the features and labels of the data. This is done after the data has been loaded. The features are the input data that the model will use for prediction, while the labels are the correct output or target values that the model is trying to predict. Collecting the features and labels is an important step in preparing the data for training. Shown below in fig 3.5 are feature collected from the dataset when it initiates the collect_features_ labels function.

Fig. 6 Function for the collection feature labels

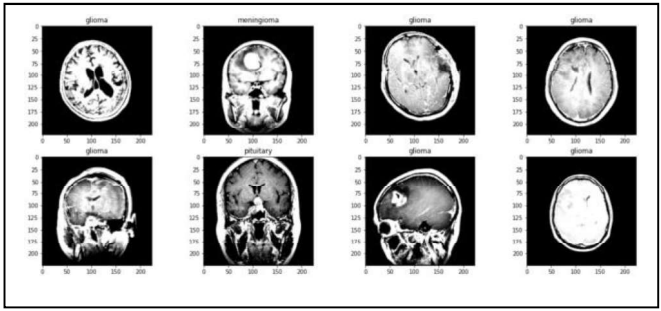

Fig. 7 Some examples of extracted features

*E. Data augmentation*

A convolutional neural network is set to be optimized when its parameter tuning can correctly map a set of input (images, numeric variables, categorical variables etc.) to a label, also known as the output, without loss to its model regardless of the orientation of the image input. One of the ways to ensure this is to train the model until invariance is ascertained. Regardless of object orientation, size or angle of perspective, the model correctly classifies the images. Fig 3.6 shows the change in orientation of the data features. In a real-life application, the dataset of images used to train a convolutional neural network may exist in limited forms, which may not look exactly like the images that may serve as input at the application end. Since it will be time demanding to manually change the orientation for each image, it is advised to augment the data. This serves both to skew the parameters and increase the set of data the model is being trained on, thus enhancing the model's performance.

In our model, we trained new samples from existing ones by applying random transformations to old samples. This is often done in order to increase the size of the training dataset and to reduce overfitting. The function shown in fig 1.2 demonstrates how the data augmentation is carried out in the model.

```
def augmentated_images(self):
    images_after_aug = []
    labels_after_aug = []
    augmentation_object = self.DataAugmentation_Object()
    for index, image in enumerate(self.images):
        for i in range(3):
            img = augmentation_object.flow(np.reshape(image, (1, 224, 224, 3)
            images_after_aug.append(np.reshape(img, (224, 224, 3)))
            labels_after_aug.append(self.labels[index])
    self.labels = np.asarray(labels_after_aug)
    self.images = np.asarray(images_after_aug)
```

```
Brain_Tumor_obj.augmentated_images()

plt.figure(figsize = (20, 8))
for i in range(8):
    plt.subplot(2, 4, i + 1)
    plt.imshow(Brain_Tumor_obj.images[i])
    plt.title(Brain_Tumor_obj.classes[Brain_Tumor_obj.labels[i] - 1])
plt.show()
```

Fig. 8 Code snippet of the python function created for data augmentation

The results of the augmented images are shown in Fig. 9

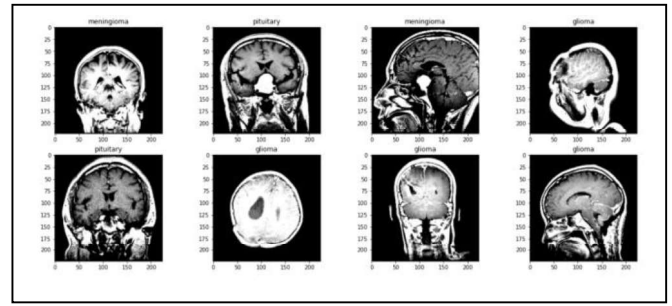

Fig. 9 Some examples of augmented images

*F. Class labels*

The code snippet shown in fig 1.3 defines a class called Brain_Tumor that has a constructor method __init__. The __init__ method is called when an object of the Brain_Tumor class is created, and it initializes the object with the provided arguments. The arguments of the __init__ method are epochs, batch_size, dataset_folder, optimizer, and loss. These arguments are assigned to instance variables of the same name, which can be accessed and modified through the object as shown in fig 1.4a,b,c.

```
class Brain_Tumor:
    def __init__(self, epochs,
                 batch_size,
                 dataset_folder,
                 optimizer,
                 loss):
        self.epochs = epochs
        self.batch_size = batch_size
        self.dataset_folder = dataset_folder
        self.optimizer = optimizer
        self.loss = loss
        self.DatasetFiles = list(pathlib.Path(os.path.join(self.dataset_fold
        self.classes = ["meningioma", "glioma","pituitary"]
```

Fig. 10 Code snippet for creating class labels.

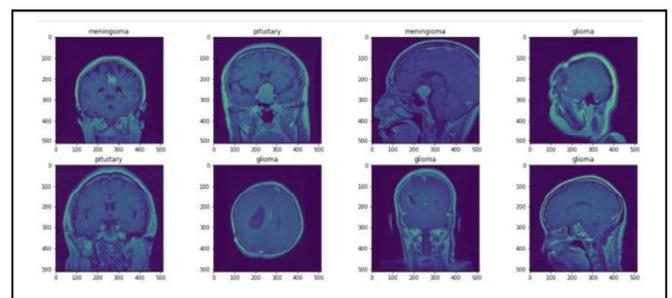

Fig. 11 Identified class labels of data features.

*G. Train-test splitting*

Splitting the dataset into training and testing sets would allow us to evaluate the performance of the model on new data. The train_test_split() calls on an object of the Brain_Tumor class. The function is used in the code to split the dataset of brain tissue images into a training set and a test set, with the test set comprising 10% of the total dataset (indicated by the 0.1 argument as shown in figure 12). This is a step in the machine learning pipelines used to evaluate the performance of the model on new data after training. The training set is used to fit the model, while the test set is used to evaluate the model's performance.

```
Brain_Tumor_obj.train_test_split(0.1)
```

Fig. 12 Train-test splitting function with an argument of 0.1

### H. CNN Architecture

CNNs are commonly used for image classification tasks. They are well-suited for this task because they can automatically learn hierarchical representations of visual data, which makes them effective at identifying patterns in images.

The Model() function defines the architecture of the neural network. It starts by creating a DenseNet121 model with no pre-trained weights and an input shape of (224, 224, 3). Then, it adds a Dropout layer with a rate of 0.8. Next, it adds a GlobalAveragePooling2D layer, which takes the average of all feature maps and flattens them into a single 1D tensor. This is followed by another Dropout layer with a rate of 0.8.

It adds a Dense layer with 3 output units, using the "GlorotNormal" initializer and the "SoftMax" activation function. It also includes L2 regularization on the kernel and bias weights to help prevent overfitting. The model's input 'd' is set to be the output of the DenseNet121 model, and the output 'm' is set to be the output of the Dense layer. The model then loads pre-trained weights from the "chexnet-weights/brucechou1983_CheXNet_Keras_0.3.0_weights.h5" file and sets the first 200 layers to be non-trainable, while allowing the remaining layers to be trainable. The model is then stored as an attribute of the Brain_Tumor object.

The compile() function then compiles the model using the specified optimizer, loss function and accuracy.

Finally, the fit_model() function trains the model on the training data (self.X_train and self.y_train) for the specified number of epochs and batch size, using the validation data (self.X_test and self.y_test) to evaluate the model's performance at each epoch. It also includes two callbacks: a ReduceLROnPlateau callback, which reduces the learning rate when the validation loss has stopped improving, and an EarlyStopping callback, which stops the training when the validation loss has not improved for a specified number of epochs. The training history is then stored as an attribute of the Brain_Tumor object.

```
def Model(self):
    d = densenet.DenseNet121(weights=None, include_top = False, input_shape = (224, 224, 3))
    m = tf.keras.layers.Dropout(0.8)(d.output)
    m = tf.keras.layers.GlobalAveragePooling2D(name = "GlobalAveragePooling2D_")(m)
    m = tf.keras.layers.Dropout(0.8)(m)
    m = tf.keras.layers.Dense(3, kernel_initializer=GlorotNormal(),
                              activation = 'softmax', kernel_regularizer= tf.keras.regularizers.l
                              bias_regularizer= tf.keras.regularizers.L2(0.0001))(m)
    m = tf.keras.models.Model(inputs = d.input, outputs = m)
    m.load_weights("chexnet-weights/brucechou1983_CheXNet_Keras_0.3.0_weights.h5", by_name=True,
    for layer in m.layers[:200]:
        layer.trainable = False
    for layer in m.layers[200:]:
        layer.trainable = True
    self.m = m
def compile(self):
    self.m.compile(optimizer = self.optimizer
                   , loss = self.loss, metrics = ['accuracy'])
```

Fig. 13 Non-hybrid CNN model

Defining and compiling the architecture of the CNN, including the number and size of the convolutional and pooling layers, as well as the activation functions and regularization techniques to be used. The Brain_Tumor_obj.Model() function initializes the model architecture and loads the pre-trained weights. The Brain_Tumor_obj.compile() function sets the optimizer, loss function, and any other metrics to be used during training. The Brain_Tumor_obj.fit_model() function trains the model on the training dataset and evaluates its performance on the validation dataset.

```
Brain_Tumor_obj.to_categorical_label()

Brain_Tumor_obj.train_test_split(0.1)

Brain_Tumor_obj.Model()
Brain_Tumor_obj.compile()
Brain_Tumor_obj.fit_model()

Epoch 1/25
4136/4136 [==============================] - 171s 36ms/step - loss: 0.9056 - accuracy
lr: 0.0010
Epoch 2/25
```

Fig. 14 Batch of input data passed through a total of 25 Epochs

### I. XGBOOST Architecture

Convolutional eXtreme Gradient Boosting consists of several stacked convolutional layers to recognize input features and is able to learn the features automatically, after performing the required training of the convolutional neural network on brain tumour images, we will extract the last layer preceding the Dense classification layers (the layer named GlobalAveragePooling2D_ while building the proposed neural network) and from it we will extract the features and train XGBoost in the last layer. The output of the CNN models is parsed as input and fitted into the xgboost framework to minimize the loss function and improve the model's accuracy.

We do this by loading the previously trained CNN model from two separate files: the model architecture ("brain_tumor_model.h5") and the model weights ("brain_tumor_weights.h5"). Then, we created a new model that takes the input and output layers of the loaded model, but only includes the layers up to and including the layer called "GlobalAveragePooling2D_". This new model is referred to as "new_model". Finally, the "predict" method of the new model generates features for the training data. The shape of the resulting feature matrix is then printed to the console.

XGBoost is a gradient boosting algorithm that is commonly used for classification tasks. It is trained on the feature maps extracted from the CNN model (X_train_features) and the corresponding ground truth labels (Brain_Tumor_obj.y_train). The model is then used to make predictions on the test set (X_test_features) and the predicted labels (y_pred) are compared to the ground truth labels (y_test) to evaluate the performance of the model. The learning rate, maximum depth of the tree, and number of estimators are specified as hyperparameters for the model. The model performs a multi-class classification task through use of the 'multiclass: Softmax' function to predict the class probabilities as shown in Fig. 15.

```
from xgboost import XGBClassifier

xgb = XGBClassifier(objective='multiclass:softmax', learning_rate = 0.1,
                    max_depth = 15, n_estimators = 500)
xgb.fit(X_train_features, np.argmax(Brain_Tumor_obj.y_train, axis = 1))

XGBClassifier(max_depth=15, n_estimators=500, objective='multi:softprob')

X_test_features = new_model.predict(Brain_Tumor_obj.X_test)

y_pred = xgb.predict(X_test_features)

y_test = np.argmax(Brain_Tumor_obj.y_test, axis =1)
```

Fig. 15 The output of the CNN models is parsed as input and fitted into the xgboost framework.

## IV. DISCUSSION AND ANALYSIS

We created a plot of the training and validation loss over the course of the training epochs as shown in figure 16. The history attribute of the Brain_Tumor object stores the training history of the model, including the loss and validation loss at each epoch. The pyplot function of the matplotlib module is used to create the plot, with the training loss and validation loss being plotted using the training loss and val_loss keys of the history dictionary, respectively. The training loss is used to evaluate the model's performance on the training set and to adjust the model's weights and biases during training. The validation loss is used to evaluate the model's generalization ability, or its ability to perform well on new data.

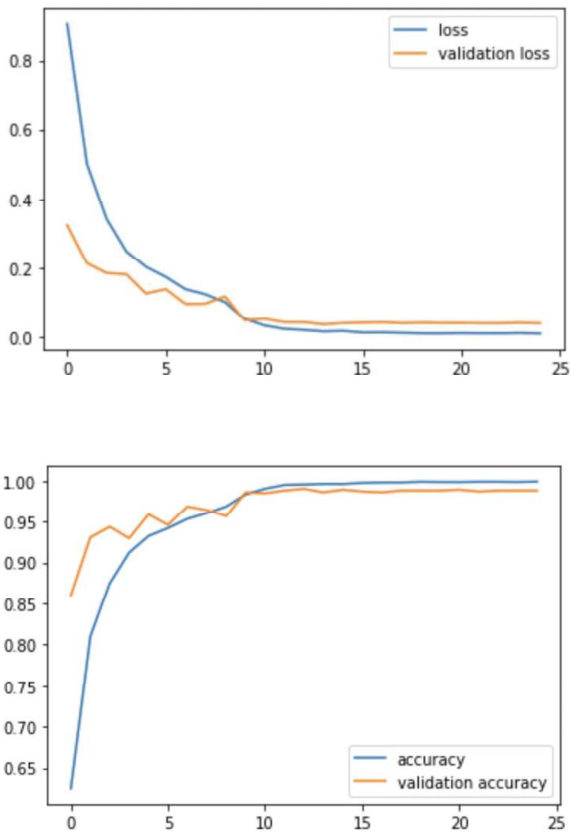

Fig. 16 Training and Validation loss and accuracy for the DenseNet20-based non-hybrid CNN model

Minimum Loss is achieved for both DenseNet201-CNN and DenseNet201-C-XGboost, whereas validation accuracy for C-XGBoost as shown in Figure 17b is slightly higher than that of purely DenseNet201 based CNN model. A validation accuracy of 0.99021739 is achieved for the proposed model while a slightly lower accuracy of 0.9880434. If the training loss is much lower than the validation loss, it could indicate that the model is overfitting, meaning it is performing well on the training data but not generalizing well to new data, in which case, the model may benefit from regularization techniques such as weight decay or dropout.

If the training loss is much higher than the validation loss, it could indicate that the model is underfitting, meaning it is not able to capture the underlying pattern in the training data well, in which case, the model may benefit from increasing the complexity of the model, such as adding more layers or increasing the number of units in each layer. For our hybrid model, the training loss and validation loss are both low and directionally close to each other. This indicates that the model is performing well and is not overfitting or underfitting.

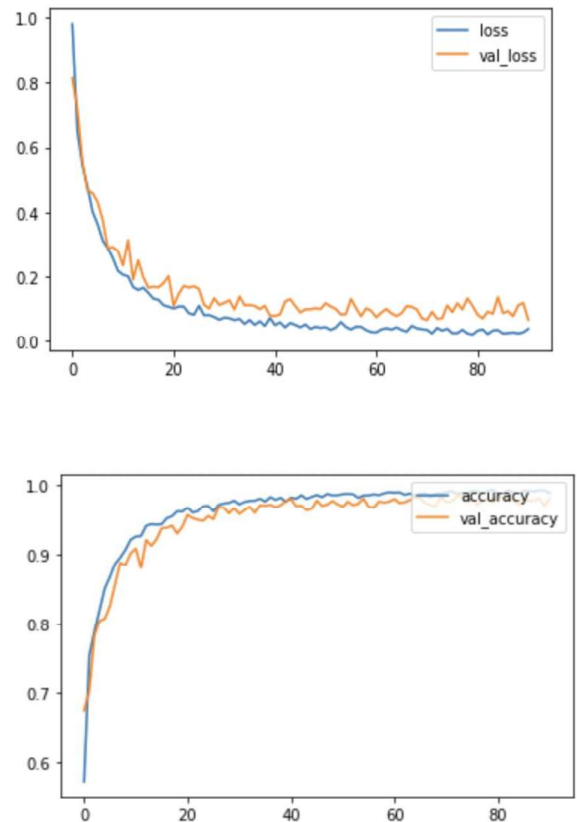

Fig. 17 Training and Validation loss and accuracy for the DenseNet201 based C-Xgboost model

### A. Evaluation Metrics

We observe and evaluate the performance of the model on a test dataset to assess its accuracy and identify any potential improvements. Evaluation metrics are used to evaluate the efficiency of an AI model, usually based on accuracy, sensitivity, specificity, F1-score, and area under the curve (AUC) of ROC charts.

Accuracy: This is a simple metric that measures correct predictions made by the model. It is computed by dividing the number of correct predictions by the total number of predictions made by the model [21]. However, accuracy can be misleading when the classes in the dataset are imbalanced, since the model can achieve high accuracy by always predicting the majority class. Hence, we computed our outcomes with other metrics such as F1 score and AUC.

Accuracy = (TP + TN) / TO

Sensitivity, also known as true positive rate, measures the percentage of positive examples that were correctly predicted by the model. It is calculated by dividing the number of true positive predictions by the total number of positive examples in the dataset. Sensitivity is useful for evaluating the model's ability to correctly identify positive examples.

Sensitivity = TP / (TP + FN)

Specificity, also known as true negative rate, measures the percentage of negative examples that were correctly predicted by the model. It is calculated by dividing the number of true negative predictions by the total number of negative examples in the dataset [22]. Specificity is useful for evaluating the model's ability to correctly identify negative examples.

Specificity = TN / (TN+ FP)

Where TP is the number of true positive examples, FN is the number of false negative examples, FP is the number of false positive examples, and TN is the number of true negative examples. TO is the number of total observations.

F1-score is a metric that combines precision and recall into a single score. It is calculated by taking the harmonic mean of precision and recall [21]. F1-score is useful for evaluating the overall performance of a model, since it considers both the number of correct predictions and the number of false positives and false negatives.

F1-score = 2 * (P * R) / (P + R)
P = precision
R = Recall

Table 2. performance characteristics of proposed C-XGBoost model.

|  | Accuracy | F1 score | Specificity | Sensitivity |
|---|---|---|---|---|
| CNN MODEL | 98.8 | 0.97 | 95.2 | 87.4 |
| C-XGBOOST MODEL | 99.02 | 0.98 | 97.4 | 91.5 |

*B. Implementation Of Evaluation Metrics for CNN Model and Hybrid CNN-XGBoost Model in Python*

1) Accuracy

The lines of code shown in Figure 18 perform prediction tasks on the test set using the trained model and calculate the accuracy of the model on the test set. The y_pred uses the predict method of the trained model to make predictions on the test set and converts the predicted probabilities for each class into class labels by taking the argmax along the axis 1. The y_test also converts the true labels for the test set into class labels in the same way. Finally, the accuracy_score function from the sklearn.metrics module is imported and used to calculate the accuracy of the model on the test set, by comparing the predicted class labels with the true labels. The accuracy is returned as a float value between 0 and 1 which in the case of our CNN model is an impressive 0.9880434.

Fig. 18 Accuracy outcome for non-hybrid CNN model.

Fig. 19 Accuracy outcome for CNN-XGBOOST model.

This study shows the difference between the accuracy that was achieved using standalone CNN model, and that of when combined with XGBoost. The result displayed in Figure 19 shows that the integration of CNN with XGBoost leads to a higher accuracy of 0.99021739 than that achieved using CNN only.

2) Confusion Matrix

Confusion matrix is a table that is often used to describe the performance of a classification model on a dataset for which the true values are known. It is called a "confusion" matrix because it allows us to visualize the predicted and actual classifications of a model, and "confuse" these predictions with the true values.

A confusion matrix typically has four entries: true positive (TP), false positive (FP), true negative (TN), and false negative (FN). These entries are calculated by comparing the predicted values of the model with the true values. For example, a true positive is an example where the model correctly predicts the positive class, while a false positive is an example where the model incorrectly predicts the positive class.

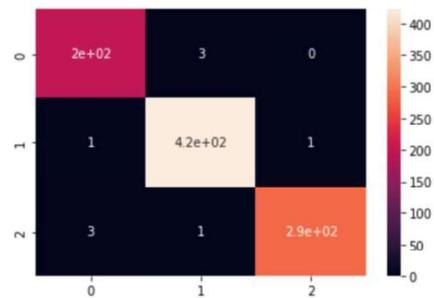

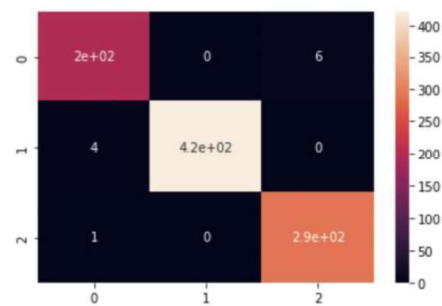

Fig. 19 Accuracy outcome for non-hybrid CNN (left) and CNN-XGBOOST model (right).

We used the seaborn library to generate a heatmap visualization of a confusion matrix for the evaluation of the performance of the model. It compares the predicted class labels (y_pred) with the true class labels (y_test) and shows the number of correct and incorrect predictions for each class. The heatmap is a visual representation of the confusion matrix, with the colour intensity indicating the number of predictions in each category. The tf.math.confusion_matrix function is used to compute the confusion matrix from the true labels (y_test) and predicted labels (y_pred). The annot parameter specifies that the values of the confusion matrix should be displayed in each cell of the heatmap. This visualization identifies which classes are being misclassified and how well the model is performing overall. The three classes represented in the matrix represents glioma - meningioma - and pituitary. The precise number of images classified by the model can be determined using the diagonal values. From the result above, we observed that the total frequency of misclassifications is fewer in C-Xgboost than purely CNN.

Random samples of 15 test images, with their predicted labels and ground truth is shown below using the proposed convolutional xgboost model.

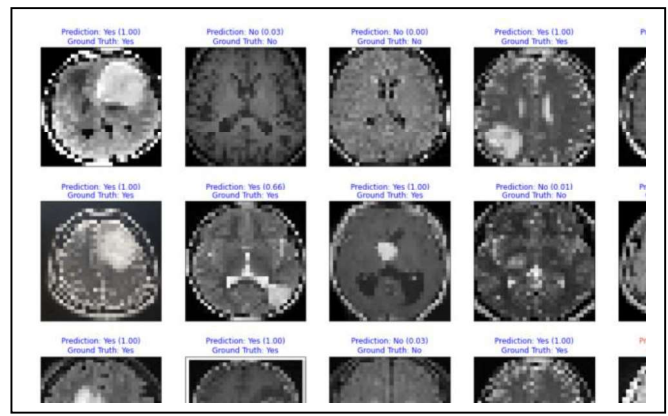

Figure 20: Random samples of 15 test images showing their predicted and ground truth labels

## C. Comparative Performance Analysis

Performance results obtained from existing deep learning models on the brain tumor MRI images are shown in comparison with that of the proposed models in Table 14. The table shows the machine learning technique employed used by the authors and their outcomes.

Table 3. performance characteristics of different prediction models trained using brain MRI images

| Reference | Study Title | Machine Learning Technique Used | Accuracy |
|---|---|---|---|
| Ahmad et al. [23] | Automatic Detection of Brain Tumors in magnetic resonance imaging (MRI) Using Machine Learning Algorithms | A combination of support vector machines, decision trees, and neural networks. | 97.50% |
| Khan et al. [24] | Brain Tumor Detection and Classification Using Multimodal MRI Scans and Deep Convolutional Neural Networks | Convolutional Neural Network | Sensitivity: 98.3% Specificity: 98.7% |
| Chen et al. [25] | Brain Tumor Detection and Segmentation in MRI Images Using Deep Learning | Deep learnig technique | Dice co-efficient: 0.82 |
| Kamnitsas et al. [26] | A Deep Learning Framework for Brain Tumor Detection and Segmentation in 3D Magnetic Resonance Images | 3D Convolutional Neural Network | Dice co-efficient: 0.81 |
| El-Shenawee et al. [27] | Convolutional Neural Networks (CNN) and adaptive thresholding | Deep Convolutional Neural Network | 95.44% |
| Al-Shara et al. [28] | Detection of Brain Tumor in MRI Images Using Deep Learning and Fast R-CNN | Deep learning and Fast R-CNN | 92.31% |
| Yan et al. [29] | Deep Learning-Based Detection of Brain Tumors in MRI Images Using U-Net. | Deep learning and U-Net. | 96.90% |
| Krishna et al. [30] | Detection of Brain Tumors in MRI Images Using a Hybrid Approach Based on Deep Learning and Transfer Learning | Deep learning and transfer learning. | 98.40% |
| Proposed Methodology | Brain tumor detection using convolutional XGBoost | CNN model | 98.80% |
| | | C-XGBOOST model | 99.02% |

## V. CONCLUSION

This paper analysed the performance of C-XGBoost models for predicting brain tumors. The C-XGBoost model which is based on DenseNet201 outperformed the non-hybrid CNN model in various metrics as discussed in the result analysis section. The dataset was acquired from the figshare public repository. The proposed hybrid C-XGBoost model leveraged the strengths of both CNNs and XGBoost, resulting in an improved ability to accurately classify brain MRIs with tumour and without tumours. The model achieved an accuracy of 99.0% and F1 score of 0.97. To assess the relative effectiveness of the proposed C-XGBoost model, its performance was also compared to that of other existing models. One of the benefits of using the XGBoost component in the proposed C-XGBoost model is its ability to provide interpretability and explainability for the model's decisions. This can be a useful diagnostic tool for medical practitioners when applied in real-time.

Further analysis and validation on larger and more diverse datasets are necessary to fully assess the capabilities of the proposed model. To further improve the model's ability to detect brain tumors at very early stages, it may be necessary to address challenges related to adjusting image capture, improving image quality, combining diverse data formats, and correcting weight misalignment. These challenges can be addressed by continually fine-tuning the model, adjusting its hyperparameters, and training it with new data.

References

[1] C. R. UK. "Cancer incidence by age." Cancer Research UK. http://www.cancerresearchuk.org/health-professional/cancer-statistics/incidence/age (accessed.

[2] Z. Qiao, J. Ge, W. He, X. Xu, and J. He, "Artificial Intelligence Algorithm-Based Computerized Tomography Image Features Combined with Serum Tumor Markers for Diagnosis of Pancreatic Cancer," (in eng), Comput Math Methods Med, vol. 2022, p. 8979404, 2022, doi: 10.1155/2022/8979404.

[3] S. Uddin, A. Khan, M. E. Hossain, and M. A. Moni, "Comparing different supervised machine learning algorithms for disease prediction," (in eng), BMC Med Inform Decis Mak, vol. 19, no. 1, p. 281, Dec 21 2019, doi: 10.1186/s12911-019-1004-8.

[4] Y. Lecun, L. Bottou, Y. Bengio, and P. Haffner, "Gradient-based learning applied to document recognition," (in English), P Ieee, vol. 86, no. 11, pp. 2278-2324, Nov 1998, doi: Doi 10.1109/5.726791.

[5] M. Gjoreski et al., "Facial EMG sensing for monitoring affect using a wearable device," (in eng), Sci Rep, vol. 12, no. 1, p. 16876, Oct 7 2022.

[6] E. A. Smirnov, D. M. Timoshenko, and S. N. Andrianov, "Comparison of Regularization Methods for ImageNet Classification with Deep Convolutional Neural Networks," (in English), Aasri Proc, vol. 6, pp. 89-94, 2014, doi: 10.1016/j.aasri.2014.05.013.

[7] P. Grohs and G. Kutyniok, Eds. Mathematical Aspects of Deep Learning. Cambridge: Cambridge University Press, 2022.

[8] A. S. El-Baz and J. S. Suri, Artificial intelligence in cancer diagnosis and prognosis. Volume 3, Brain and prostate cancer. Bristol [England] (No.2 The Distillery, Glassfields, Avon Street, Bristol, BS2 0GR, UK): IOP Publishing, 2022.

[9] J. Brownlee, XGBoost With Python: Gradient Boosted Trees with XGBoost and scikit-learn. Machine Learning Mastery, 2016.

[10] A. L. Ross Russell et al., "Spectrum, risk factors and outcomes of neurological and psychiatric complications of COVID-19: a UK-wide cross-sectional surveillance study," Brain Communications, vol. 3, no. 3, 2021, doi: 10.1093/braincomms/fcab168.

[11] G. Alfonso Perez and J. Caballero Villarraso, "Neural Network Aided Detection of Huntington Disease," (in eng), J Clin Med, vol. 11, no. 8, Apr 10 2022, doi: 10.3390/jcm11082110.

[12] J. Zhou, Q. Zhang, and B. Zhang, "Two-phase non-invasive multi-disease detection via sublingual region," (in eng), Comput Biol Med, vol. 137, p. 104782, Oct 2021, doi: 10.1016/j.compbiomed.2021.104782.

[13] H. Koo and J. Kim, "Deep learning-based diagnosis of breast cancer using mammography images," in 2018 26th International Conference on Pattern Recognition (ICPR), 2018, pp. 1236–1241.

[14] M. Ezzat, A. M. Elshahat, K. A. Elsharkawy, M. Zaky, and M. M. El-Baz, "Predicting the onset and progression of Alzheimer's disease using machine learning," in 2018 10th International Conference on Machine Learning and Computing (ICMLC), 2018, pp. 277–281.

[15] M. P. George, B. Agrawal, and P. J. George, "Predicting diabetes using logistic regression," in 2017 2nd International Conference on Computer Science and Engineering, 2017, pp. 1–4.

[16] S. P. Mavroforakis, G. A. Vouros, and I. Maglogiannis, "A machine learning approach for early diagnosis of Parkinson's disease using non-invasive biomarkers," in 2017 39th Annual International Conference of the IEEE Engineering in Medicine and Biology Society, 2017, pp. 842–845.

[17] M. A. Imran, M. Bennamoun, and R. Owens, "A deep learning approach for predicting the severity of diabetic retinopathy," in 2017 39th Annual International Conference of the IEEE Engineering in Medicine and Biology Society, 2017, pp. 4596–4599.

[18] M. G. Pechenizkiy, T. P. Kersting, J. F. F. Ziegler, and J. Gama, "Predicting heart failure using decision tree and random forest algorithms," in 2015 37th Annual International Conference of the IEEE Engineering in Medicine and Biology Society, 2015.

[19] S. Gao, L. Chen, J. Wang, and X. Li, "Predicting the likelihood of stroke using a hybrid machine learning approach," in 2014 36th Annual International Conference of the IEEE Engineering in Medicine and Biology Society, 2014, pp. 5580–5583.

[20] Y. Wang, X. Chen, and J. Li, "Predicting the likelihood of breast cancer using support vector machine," in 2012 34th Annual International Conference of the IEEE Engineering in Medicine and Biology Society (EMBC), 2012, pp. 4644–4647.

[21] Kotu, N., & Alahakoon, D. (2018). Classification model performance evaluation metrics. In 2018 IEEE 9th International Conference on Software Quality, Reliability, and Security Companion (QRS-C) (pp. 109-115). IEEE.

[22] Hossain, M. J., & Rahman, M. M. (2015). Performance evaluation of classification models. International Journal of Computer Applications, 111(7)